# Anomalous excitonic resonance Raman effects in few-layer $MoS_2$

*Jae-Ung Lee,[a,‡] Jaesung Park,[a,‡] Young-Woo Son[b] and Hyeonsik Cheong[a,]**

[a] Department of Physics, Sogang University, Seoul 121-742, Korea.

[b] School of Computational Sciences, Korean Institute for Advanced Study, Seoul 130-722, Korea

**CORRESPONDING AUTHOR FOOTNOTE:** E-mail: hcheong@sogang.ac.kr

**ABSTRACT:** The resonance effects on the Raman spectra from 5 to 900 $cm^{-1}$ of few-layer $MoS_2$ thin films up to 14-layers were investigated by using six excitation energies. For the main first-order Raman peaks, the intensity maximum occurs at ~2.8 eV for single layer and at ~2.5 eV for few-layer $MoS_2$, which correspond to the band-gap energy. At the excitation energy of 1.96 eV, several anomalous behaviors are observed. Many second-order peaks are anomalously enhanced even though the main first-order peaks are not enhanced. In the low-frequency region (<100 $cm^{-1}$), a broad peak centered at ~38 $cm^{-1}$ and its second order peak at 76 $cm^{-1}$ appear for the excitation energy of 1.96 eV. These anomalous resonance effects are interpreted as being due to strong resonance with excitons or exciton-polaritons.

## 1. Introduction

Transition metal dichalcogenides (TMDCs) such as $MoS_2$ and $WSe_2$ films are attracting much interest as semiconducting 2-dimensional (2D) materials for flexible electronic devices.[1,2] In addition to possible applications, these materials are intensely investigated as model systems for studying electronic interactions in ideally confined 2D systems in comparison with semiconductor quantum wells which are only quasi-2D. Of the TMDCs, $MoS_2$ is the most extensively studied.[2–4] Owing to the band gap which is tunable by controlling the thickness, $MoS_2$ is expected to complement or replace graphene in electronic devices such as field effect transistors[1,5,6] or photodetectors[6–8] where a sizable band gap is necessary. The optical properties of $MoS_2$ are also attracting interest. Strong photoluminescence (PL) signal from excitonic states are observed at all thicknesses.[2,9] A negative trion state, where two electrons and a hole form a bound state, has also been observed.[10]

Single-layer $MoS_2$ can have either trigonal prism (1H-$MoS_2$) or octahedral (1T-$MoS_2$) coordination.[11,12] 1T-$MoS_2$ is metastable, and only the trigonal phase is found in natural bulk $MoS_2$.[12,13] In bulk $MoS_2$, two types of stacking, hexagonal 2H and rhombohedral 3R, are found. Naturally occurring $MoS_2$ crystals (molybdenite) are predominantly 2H type.[14] The basic building block of 2H type $MoS_2$ is

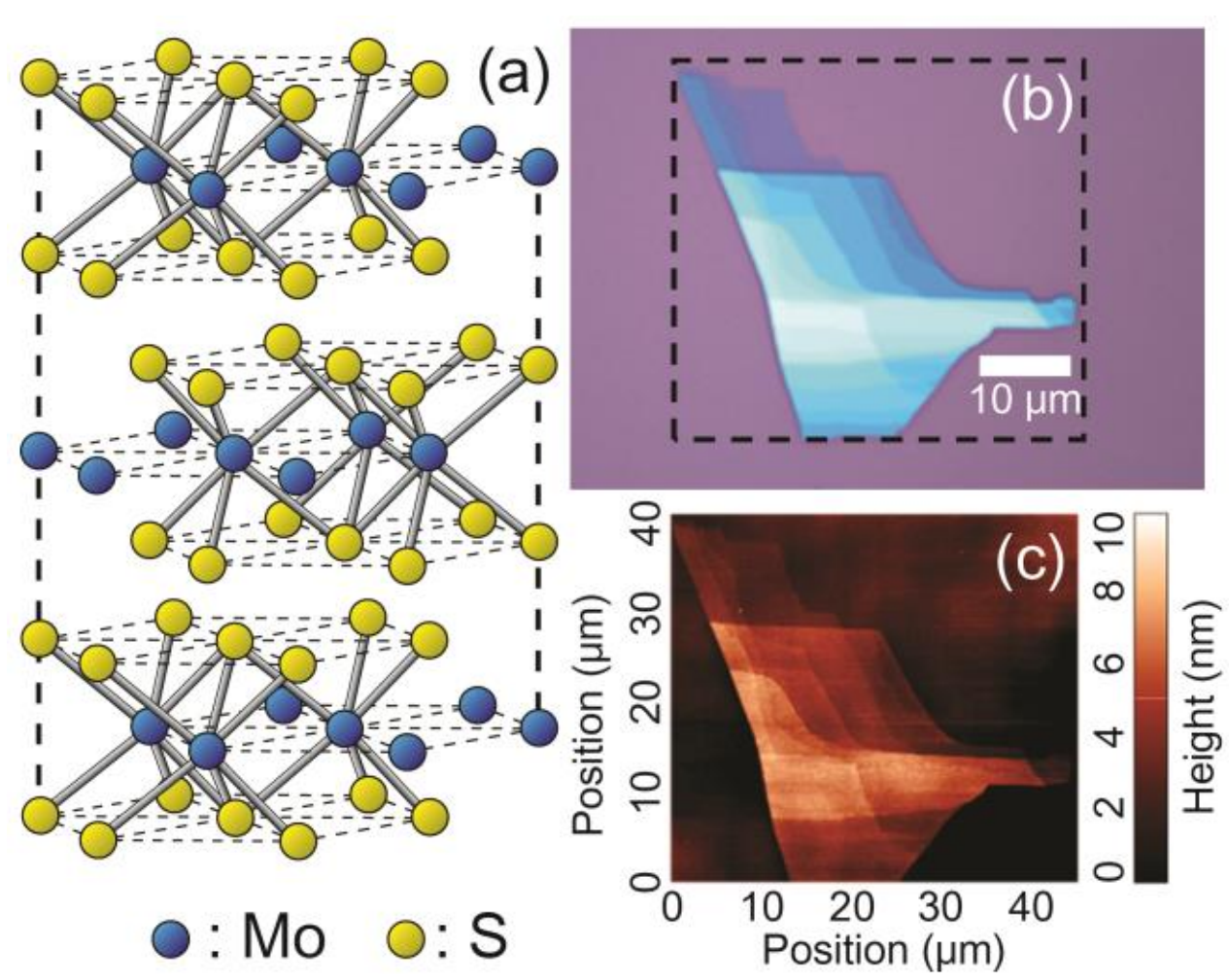

Fig. 1 (a) Crystal structure of 2H-$MoS_2$. (b) Optical microscope and (c) atomic force microscope (AFM) image of a $MoS_2$ sample on a $SiO_2$/Si substrate.

composed of two S layers and one Mo layer arranged in hexagonal lattices.[11,15] These layers are covalently bonded to one another and form a ‘trilayer’ (TL). Each TL is connected to neighboring TLs via weak van der Waals interactions.[16] The schematic of crystal structure is shown in Figure 1(a). Due to the weak coupling between the TLs, it can be easily cleaved by using the well-developed mechanical exfoliation technique.

As in the case of graphene, Raman spectroscopy has emerged as a powerful tool to measure the thickness of $MoS_2$ thin films. The separation between the two most prominent peaks $E_{2g}^{1}$ and $A_{1g}$ depends on the thickness and is widely used to determine the number of TLs.[17] In the low-frequency region (<100 $cm^{-1}$), interlayer shear and breathing modes are observed as a function of the thickness[18,19] and provide valuable information about the interlayer interactions in layered materials. Several groups have also reported the resonance effects in Raman scattering of bulk[20–23] and thin-film[24,25] $MoS_2$. The resonance effects are mostly observed when the excitation energy matches with the energy of the excitonic peaks in the PL spectra. Several weak features are enhances near the resonances.[20–25] The resonance effects on the low-frequency shear and breathing modes have not yet been studied as much, probably due to experimental difficulties. Here we report on Raman scattering measurements on $MoS_2$ thin films up to 14 TLs and the resonance effects including low frequency shear and breathing modes. Several major effects that were not observed or overlooked in previous studies are observed and interpreted in terms of anomalous resonance effects.

## 2. Experimental

The samples were prepared directly on $SiO_2$/Si substrates by mechanical exfoliation from single-crystal bulk $MoS_2$ flakes (SPI supplies). We measured several samples with different thicknesses, but the results are essentially identical. The results reported here are mostly from the sample in Fig. 1(b), which eliminates possible sample-to-sample variations. The number of TLs was determined by the combination of AFM, Raman, and PL measurements. We used 6 different excitation sources: the 325 and 441.6-nm (3.81 and 2.81 eV) lines of a He-Cd laser, the 488 and 514.5-nm (2.54 and 2.41 eV) lines of an Ar ion laser, the 532-nm (2.33 eV) line of a diode-pumped-solid-state laser, and the 632.8-nm (1.96 eV) line of a He-Ne laser. The laser beam was focused onto the sample by a 50× microscope objective lens (0.8 N.A.), and the scattered light was collected and collimated by the same objective. For measurements with the 325-nm excitation, a uv objective lens (0.5 N.A.) was used. The scattered signal was dispersed with a Jobin-Yvon Horiba iHR550 spectrometer (2400 grooves/mm) and detected with a liquid-nitrogen-cooled back-illuminated charge-coupled-device (CCD) detector. To access the low-frequency range below 100 $cm^{-1}$, reflective volume holographic filters (Ondax for 514.5-nm excitation, OptiGrate for 632.8-nm excitation) were used to reject the Rayleigh-scattered light. The laser power was kept below 0.2 mW for all of the measurements in order to avoid heating. The spectral resolution and the repeatability were about 1 $cm^{-1}$. The AFM measurements are performed by using a commercial system (NT-MDT NTEGRA Spectra).

## 3. Results and discussion

### 3.1 Resonance effects on main first-order peaks

Figure 1(b) shows the optical image of a sample which includes 1 to 14-TL areas determined by using atomic force microscopy (AFM) as shown in Fig. 1(c). We measured several samples with different thicknesses, but the results reported here are mostly from the sample in Fig. 1(b), which eliminates possible sample-to-sample variations. The PL spectra taken with the excitation energy of 2.41 eV show two peaks between 1.8 and 2.0 eV for all the thicknesses (See Fig. S1†). These peaks correspond to the A and B excitons at the K ($K'$) point, respectively. The A exciton peak exhibits a slight asymmetry, which might be due to the negatively charged trion state ($A^-$).[10] The peak positions do not vary much with the number of TLs. A weaker peak is observed at 1.3~1.5 eV except for single-layer $MoS_2$. This peak redshifts with the number of TLs and is assigned to the indirect band gap transition. These results are consistent with previous reports.[2,9] According to calculations,[3,4,26–28] single-layer $MoS_2$ is a direct-gap semiconductor with the band gap at the K point of the Brillouin zone. For 2TL or thicker, the band gap becomes indirect, with the valence band maximum at the Γ point. From the peak position of indirect-gap PL, one can determine the number of TLs up to 4 layers. Also, the intensity map of either A or B exciton peak can be used to visualize the number of TLs (Fig. S1†). However, there is still quantitative disagreement regarding the band gap energy. Although excitonic peaks are observed below 2 eV, recent calculations predict the band gap energy of single-layer $MoS_2$ to be ~2.8 eV.[4] The difference is usually attributed to the large exciton binding energy due to reduced screening in 2D materials. Our resonance Raman study may shed some light on this controversy.

Figure 2 shows the Raman spectra measured with 6 different excitation energies from a 7TL area of the sample (see also Figs. S2–S5†). To compare the absolute intensity as a function of the excitation energy, the spectra are normalized by the intensity of the first-order Raman peak from the Si substrate (520 $cm^{-1}$). The well-defined $1/\lambda^4$ dependence of the scattering intensity, the resonance effect of Si,[29] and the interference effect[30,31] from the $SiO_2$ dielectric layer are all included in the normalization process. Since the symmetry groups for bulk $MoS_2$ and few-layer $MoS_2$ are different, the peak notations should in principle be different even though they originate from similar vibrational modes.[18,19] To avoid confusion, however, we will follow the peak notations for bulk $MoS_2$. The most prominent Raman peaks are $E_{2g}^{1}$ and $A_{1g}$ modes. As the number of TLs increases, the $E_{2g}^{1}$ mode redshifts whereas the $A_{1g}$ mode blueshifts. Since these peaks are from first-order Raman scattering processes, there is no excitation energy dependence (see Fig. S6†).[32] The separation between these two peaks increase with thickness and is routinely used to identify the number of TLs.[17] Upon close inspection, one finds several weak peaks in the Raman spectra. The relative intensities of the weaker peaks depend on the excitation energy. Some peaks appear only for a specific excitation energy, suggesting a resonance effect. The peaks are assigned by comparing their positions with the phonon dispersion of bulk $MoS_2$.[20,33,34] Some peaks are assigned to first-order scattering of zero-momentum, zone-center (Γ) phonons. Others are assigned to second-order scattering involving phonons at the zone boundary (M) or at the zone center (Γ). Table 1 summarizes the peak positions and the assignments of these weaker peaks.

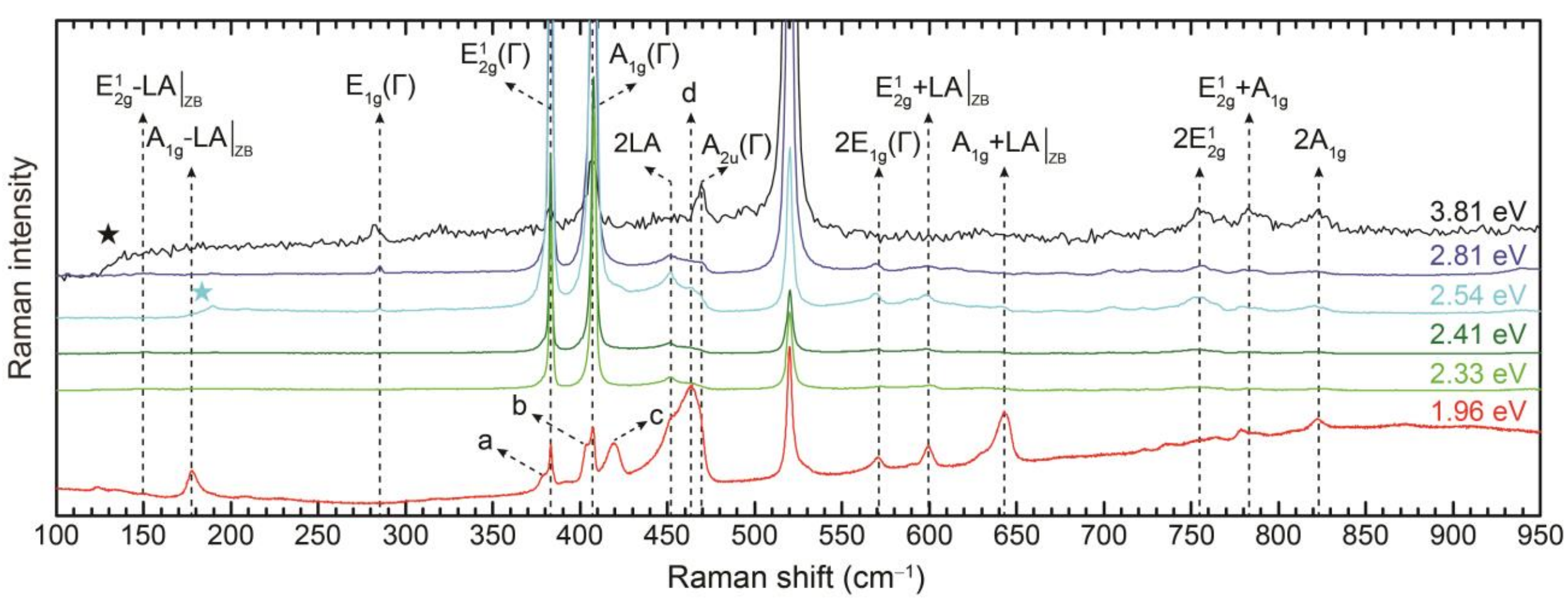

Fig. 2 Raman spectra of 7TL $MoS_2$ measured with 6 different excitation energies. The peak assignments follow the bulk case in the literature. Features indicated by (★) are experimental artefacts due to the cutoff of the edge filters.

Table 1 Raman peak positions of 7TL $MoS_2$ with 6 different excitation energies shown in Fig. 2. The units are $cm^{-1}$.

| Excitation (eV) | 3.81 | 2.81 | 2.54 | 2.41 | 2.33 | 1.96 |
|---|---|---|---|---|---|---|
| $E_{2g}^1 - LA\vert_{ZB}$ | | 150.0 | | 150.3 | 150.1 | 150.3 |
| $A_{1g} - LA\vert_{ZB}$ | | 179 | | 178.0 | 178.0 | 177.9 |
| $E_{1g}(\Gamma)$ | 283 | 284.4 | 284.7 | | | |
| a | | | | | | 376.7 |
| $E_{2g}^1(\Gamma)$ | 383.0 | 383.1 | 383.0 | 383.1 | 383.0 | 383.2 |
| b | | | | | | 403.4 |
| $A_{1g}(\Gamma)$ | 406.8 | 407.5 | 407.5 | 407.7 | 407.6 | 407.3 |
| c | | | | | | 419.1 |
| $2LA$ | | 450.7 | 450.7 | 450.7 | 450.5 | 451.3 |
| d | | | | | | 463.5 |
| $A_{2u}(\Gamma)$ | 469.3 | 469.8 | 469.3 | | | |
| $2E_{1g}(\Gamma)$ | | 568.8 | 568.5 | 568.6 | 569.0 | 571.2 |
| $E_{2g}^1 + LA\vert_{ZB}$ | | 599.5 | 598.0 | 598.0 | 599.7 | 599.0 |
| $A_{1g} + LA\vert_{ZB}$ | | | | | | 643.5 |
| $2E_{2g}^1$ | 756.9 | 756.8 | 753.5 | 753.0 | 753.7 | 754.9 |
| $E_{2g}^1 + A_{1g}$ | 782.0 | 780.3 | 780.6 | 780.5 | 782.7 | 779 |
| $2A_{1g}$ | 821.9 | 821.0 | 820.8 | 821.1 | 823 | 822.3 |

Figure 3(a) shows the resonance effect on the main Raman modes ($E^1_{2g}$ and $A_{1g}$) of 7TL (see also Figs. S7† and S8†). The intensity vs excitation energy plots in Figs. 3(b) and (c) for $E^1_{2g}$ and $A_{1g}$, respectively, show resonance effects for several TLs. For 1TL, the resonance occurs at 2.8 eV or higher. For 2TL, it is between 2.54 eV and 2.81 eV. For thicker layers, the resonance is close to 2.54 eV. Theoretical calculations show that the direct gap at the K point decrease from 1TL to 2TL, and the bulk has even smaller value.[3,26,27] Our results indicate that the direct gap does not change appreciably between 3TL and bulk. Furthermore, our results corroborate recent calculations that predict direct band gaps near 2.8 eV or higher for 1 and 2TL.[3,4]

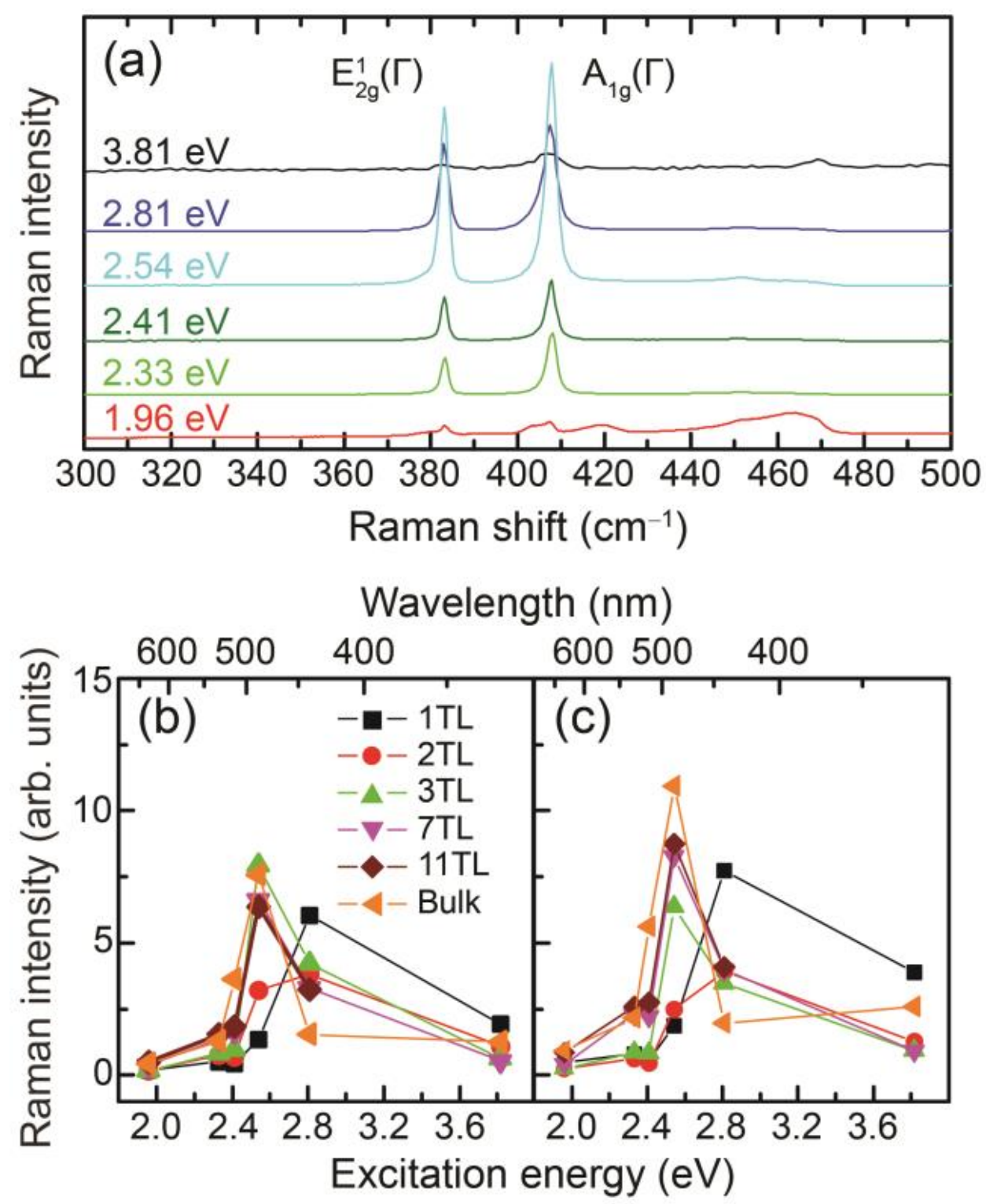

Fig. 3 (a) Two main Raman modes ($E^1_{2g}$ and $A_{1g}$) of 7TL $MoS_2$ with 6 excitation energies. Raman intensities of (b) and (c) as a function of the excitation energy for several TLs.

### 3.2 Resonance effects on forbidden modes

Figure 4 summarizes Raman spectra of 1–7TL and bulk samples measured with 6 different excitation energies (see also Fig. S9†). Some of the weaker peaks are strongly enhanced depending on the excitation energy. The sharp peak at 285 cm$^{-1}$ is observed in all samples and is strongly enhanced for excitation energies larger than 2.5 eV. This peak was previously observed in bulk $MoS_2$ and was assigned as the $E_{1g}$ mode, in which the Mo atoms are stationary and the S atoms above and below the Mo atom move in opposite directions on the lattice plane. However, the $E_{1g}$ mode in *bulk* $MoS_2$ is forbidden in backscattering geometry. In order to test the possibility that its appearance is due to oblique incidence of the excitation laser because of the large N.A. of the objective lens, we repeated the measurement with another objective lens with a smaller N.A. of 0.4. The relative intensity of this peak with respect to the main $E_{2g}^1$ peak did not change. The $E_{2g}^1$ mode in bulk $MoS_2$ corresponds to $E''$ in 1TL (or odd number of TLs) and $E_g$ in 2TL (or even number of TLs). Group theoretical considerations[18] predict that the $E''$ mode is forbidden and the $E_g$ mode is allowed in backscattering geometry. However, our data show that there is no considerable difference in the intensity of this peak between odd and even TLs. Although we tentatively label this peak as $E_{1g}(\Gamma)$ because no other phonon mode near this frequency is predicted for 1TL,[35,36] why this peak appears in backscattering Raman spectra is far from understood at this point. It should be noted that in 1TL, this peak appears clearly only for the highest excitation energy, which might imply selection rule violation triggered by resonance with higher energy band gaps.

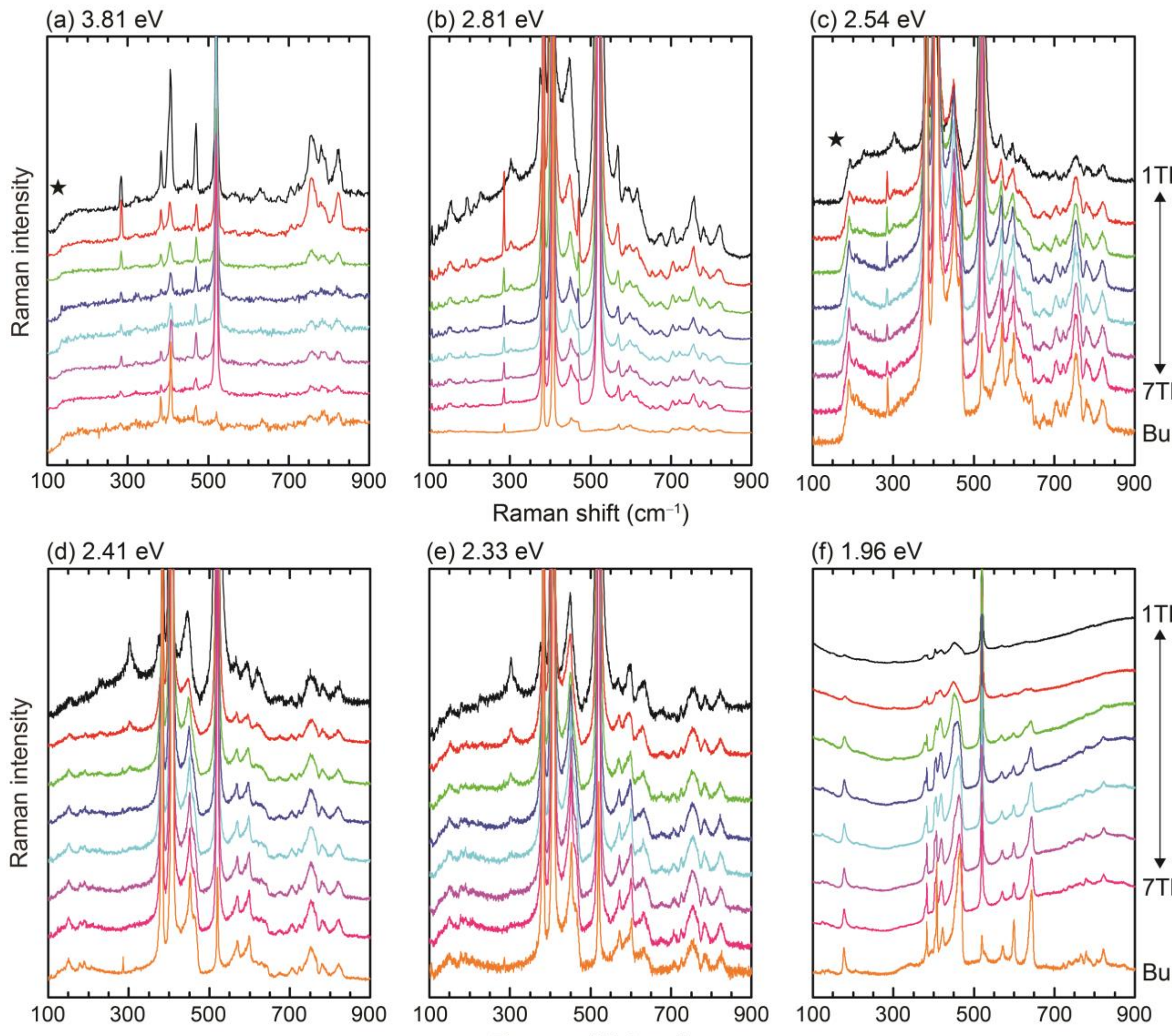

Fig. 4 Raman spectra as functions of thickness for excitation energies of (a) 3.81, (b) 2.81, (c) 2.54, (d) 2.41, (e) 2.33, and (d) 1.96 eV. Features indicated by (★) are experimental artefacts due to the cutoff of the edge filters.

The peak at 469 $cm^{-1}$ shows a trend that is very similar to the $E_{1g}(\Gamma)$ peak at 285 $cm^{-1}$. This peak is well resolved as an isolated peak for the 3.81-eV excitation. For lower excitation energies, it appears as a shoulder. Its intensity directly correlates with that of the $E_{1g}(\Gamma)$ peak. According to calculations for bulk $MoS_2$, $A_{2u}$ and $B_{2g}^1$ modes have frequencies near this peak.[35,36] These modes involve opposite vibrations of Mo and S atoms in the out-of-plane direction. For 1TL, the corresponding mode is $A_2''$, which is not Raman active. We can infer, from the excitation energy dependence, that the mechanism for the appearance of this peak, which we tentatively label as $A_{2u}(\Gamma)$, is directly related with the mechanism for the $E_{1g}(\Gamma)$ peak at 285 $cm^{-1}$.

### 3.3 Low-frequency modes

Breathing and shear modes[18,19,37] are observed in the low frequency region (<100 $cm^{-1}$). Figures 5 and S10† show these modes for Stokes and anti-Stokes scattering. The shear modes due to rigid in-plane vibrations of the layers with respect to each other blueshift with the number of TLs. The breathing modes show an opposite behaviour: they redshift with the number of TLs. These modes are quite useful for determining the number of TLs up to the largest number of TLs measured (14TL). Since these modes are from first order Raman processes, these features do not depend on the excitation energy [Fig. 5(c)].

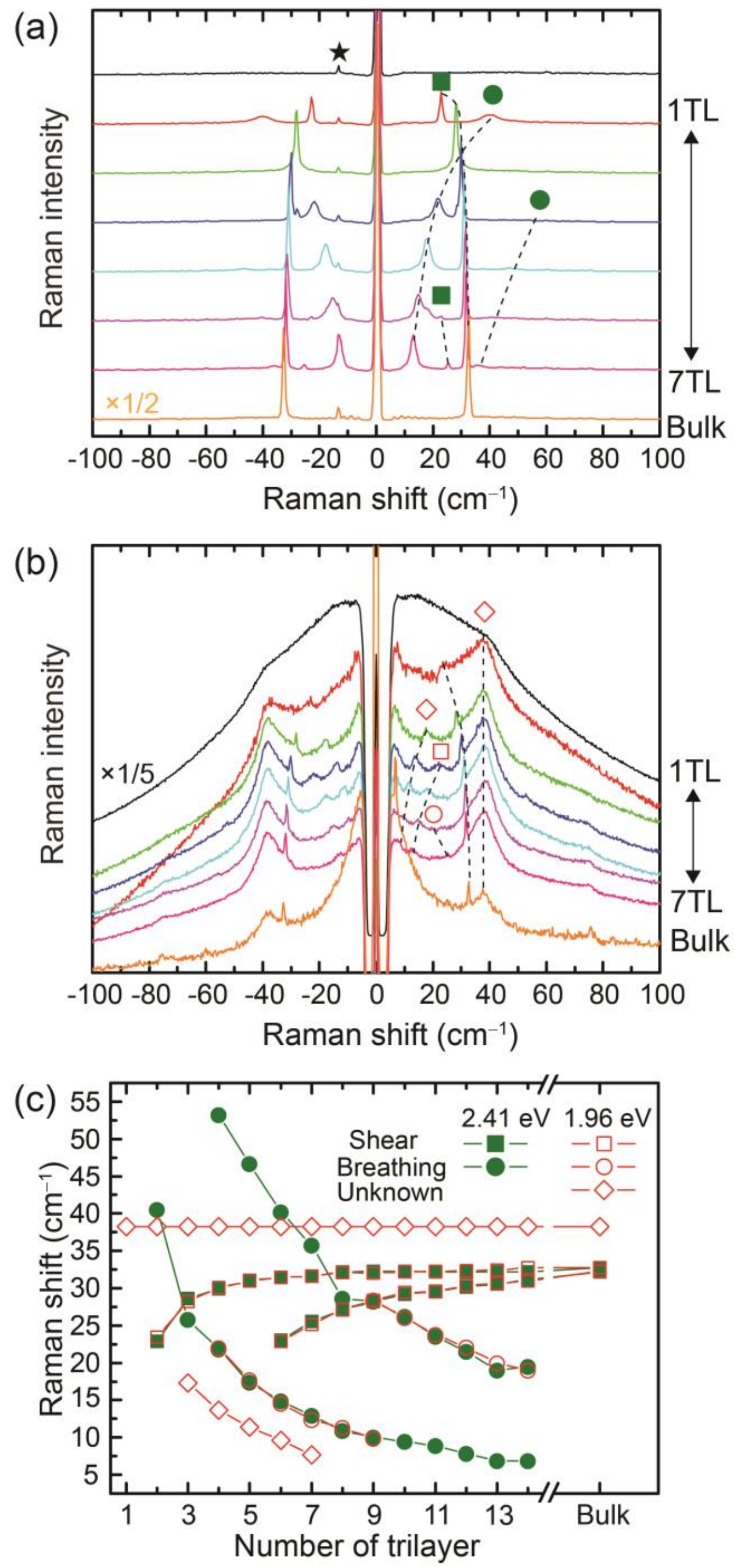

Fig. 5 Low-frequency (<100 cm$^{-1}$) Raman spectra taken with excitation energies of (a) 2.41 eV and (b) 1.96 eV. A laser plasma line is indicated by (★). In (b), the intensity for 1TL is multiplied by 1/5. (c) Peak positions for breathing (circle) and shear (square) modes as functions of the number of TLs. The features that are not explained are indicated by diamonds.

On the other hand, dramatic changes are observed for the 1.96-eV excitation, which is close to the energies of the excitonic PL peaks in Fig. S1(a)†. A broad and strong peak ($X$) centered at 38 $cm^{-1}$ and its second-order ($2X$) signal at 76 $cm^{-1}$ is seen only for the 1.96-eV excitation. The peak position does not depend on the number of TLs. It is seen even for the 1TL sample. This peak was observed before and was interpreted as an electronic Raman scattering associated with the spin-orbit coupling induced splitting in a conduction band at K points.[38] However, our circularly polarized Raman scattering measurements indicate that this peak is not due to a spin-flip scattering process [Fig. S11(a)†]. Since both the electronic band structure and the phonon dispersion depend on the number of TLs, the origin of this peak is puzzling. The fact that this peak is only observed for the 1.96-eV excitation clearly indicates that it should be related with resonance with excitons. A possible explanation is scattering with biexcitons. However, increasing the excitation intensity did not enhance this peak with respect to the shear mode signals, which essentially rules out biexcitons as a possible explanation. Further experimental and theoretical investigations are needed to identify its origin. Another peak with a frequency lower than the lowest-frequency breathing mode (13 $cm^{-1}$ for 4TL) is clearly observed for 3 to 7 TLs. This peak has not been reported in the literature. Theoretical calculations[18,19] do not predict any vibrational mode in this frequency range. It is interesting that this peak follows the trend of the breathing modes as a function of the number of TLs. Further detailed study is needed to elucidate the origin of this peak.

Another striking feature is the broad signal centered at the laser frequency. This Rayleigh-like feature is not due to stray laser light or other experimental artefacts: this feature is not strongly reduced in the cross-polarization configuration [Fig. S11(b)†]. Circularly polarized Raman scattering measurements show that this feature is due to spin-conserving scattering processes [Fig. S11(a)†]. We suggest that this ‘central peak’ is due to resonant scattering via the exciton states through emission of acoustic phonons.

Suppose that the excitation laser energy is slightly larger (smaller) than the exciton energy. An exciton can be formed by absorbing the photon and emitting (absorbing) acoustic phonons which would make up the small energy difference between the photon and the exciton. The re-emitted photon through the exciton state would be slightly shifted from the laser wavelength by the phonon energy. The momentum conservation condition is partially relaxed due to the strong binding between the electron and the hole in an exciton. Such scattering probability would decrease as the energy (and the momentum) of the phonon

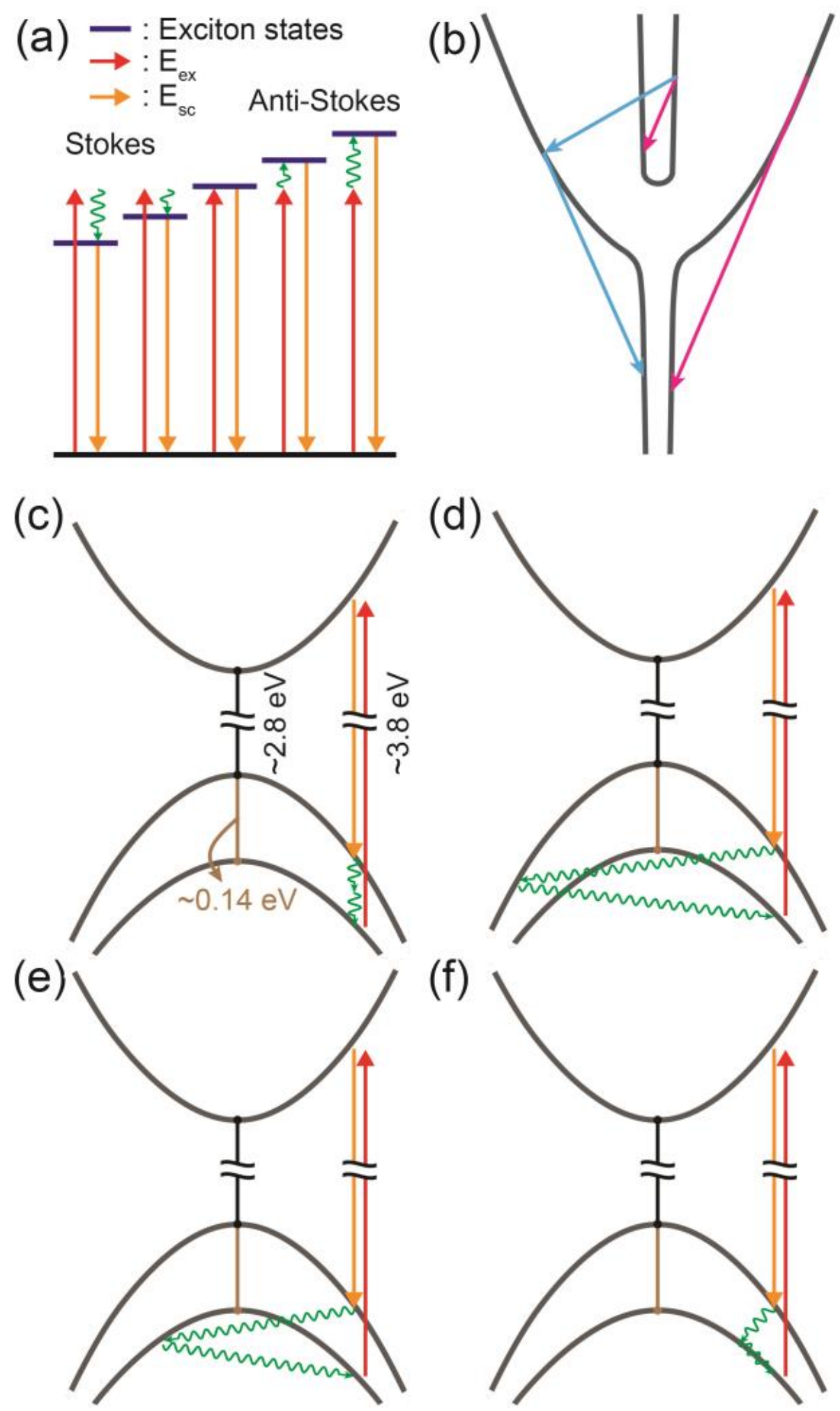

Fig. 6 (a) Schematic for the mechanism of the 'central peak' in the low-frequency spectra taken with the 1.96-eV excitation [Fig. 5(b)]. The red arrows indicate the excitation laser and the orange arrows indicate emitted photons. (b) Schematics of resonant Raman scattering processes mediated by exciton-polariton bands. The blue (magenta) arrows indicate transition of exciton-polaritons due to two-phonon (one-phonon) scattering. (c–f) Possible 'triple- or double-resonance' scattering processes for two-phonon Raman peaks in the range of 620 – 830 $cm^{-1}$.

increases. Since the exciton energy varies due to inhomogeneity in the sample or the environmental factors, the emitted (absorbed) photon energy would be centered around the excitation energy [See Fig. 6(a)].[39] Because the Stokes scattering is always stronger than the anti-Stokes scattering, the central peak is slightly asymmetric.

### 3.4 Anomalous resonance effects for 1.96-eV excitation

For the 1.96-eV excitation, other dramatic changes in the spectra are also observed [Figs. 2, 4(f) and S2–S5†]. It is surprising that some of the weaker peaks for other excitation energies are even stronger than the main $E_{2g}^{1}$ and $A_{1g}$ peaks, which may be explained in terms of resonance with the excitons. The most prominent enhancements are observed for the peak *c* at ~418 cm$^{-1}$, a broad asymmetric peak at ~460 cm$^{-1}$, and a series of peaks in the 570–650 cm$^{-1}$ range. The peak *c* is observed for all thicknesses, and its position does not show much dependence on the thickness. In *bulk* $MoS_2$, this peak was interpreted as being due to a resonant two-phonon Raman process of the successive emission of a dispersive longitudinal quasi-acoustic (QA) phonon (the breathing mode branch) and a dispersionless $E_{1u}^{2}$ phonon, both along the *c* axis,[22,40] through resonance with exciton-polariton bands near the zone center. However, the QA phonon frequency should depend on the number of TLs, which contradicts our observation. Furthermore, the appearance of this peak in the spectrum for 1TL [Figs. 4(f) and S2†] cannot be explained since a QA phonon mode in the c axis direction does not exist for 1TL. We propose that this peak is due to a scattering by the combination of the $E_{1u}^{2}$ mode (~380 cm$^{-1}$) and the *X* mode (38 cm$^{-1}$) mediated by the exciton-polariton bands [blue arrows in Fig. 6(b)]. The Raman inactive $E_{1u}^{2}$ mode may be activated due to the resonance.[41] This interpretation is further supported by the comparison of Stokes and anti-Stokes scattering data. Figure S12† compares Stokes and anti-Stokes spectra for several TLs, excited with the 1.96-eV laser line. Only this peak shows clear displacement between Stokes and anti-Stokes scattering. Because the resonance conditions for the Stokes and anti-Stokes scattering processes between exciton-polariton bands are met by modes with different momenta, the Raman shifts are different for the two cases.[40] This interpretation implies that the *X* mode is dispersive because $E_{1u}^2$ is dispersionless, which should be verified in future work.

The broad and asymmetric peak at ~460 cm$^{-1}$ is a convolution of several components. In addition to 2LA and $A_{2u}(\Gamma)$, a peak labelled *d* exists. Unlike 2LA and $A_{2u}(\Gamma)$, the peak *d* is only seen for the 1.96-eV excitation, which suggests a strong resonance effect. We interpret this peak as being due to scattering by a phonon in the $A_{2u}(\Gamma)$ branch with a finite momentum away from the Γ point, mediated by resonance with exciton-polariton transitions as schematically explained in Fig. 6(b) (magenta arrows). The dispersion of the $A_{2u}(\Gamma)$ branch is such that the phonon energy decreases slightly as the momentum moves away from the Γ point, which explains why *d* has a frequency slightly smaller than the $A_{2u}(\Gamma)$ peak. In resonance Raman of CdS,[42] a similar phenomenon has been observed. The weak side peaks labelled *a* and *b* can be explained similarly, as being due to $k \neq 0$ phonons of the $E_{2g}^1$ and $A_{1g}$ branches, mediated by an exciton-polariton transition. We note that Gołasa *et al.*[43,44] interpreted this peak *d* as being due to a combination of $E_{1g}(\mathrm{M})$ and an acoustic phonon at the M point, but in that interpretation, the selective enhancement of this peak at 1.96 eV cannot be explained.

The series of peaks in the 570–650 cm$^{-1}$ range are assigned to two-phonon scattering processes of various combinations. These modes have been observed in bulk[20,21,23] and few-layer[24,25] $MoS_2$. The combinations that include acoustic phonons have been interpreted as being due to zone-boundary phonons near the M point. However, it is not obvious why only M point phonons are involved in two-phonon scattering because the momentum conservation condition can be satisfied by any pair of phonons with opposite momenta. For the $A_{1g} + LA\big|_{ZB}$ peak, for example, the intensity would be maximum if $A_{1g}(\vec{q})$ and $LA(-\vec{q})$ have large densities of states (DOS's) for a certain $\vec{q}$. The optical phonon $A_{1g}$ has a relatively flat dispersion (large DOS) whereas the acoustic phonon LA has a large

DOS only near the zone boundaries (M or K). Since the frequencies of these phonon branches near the M and K points are more or less similar,[35,36] $A_{1g}(\mathrm{M})+LA(\mathrm{M})$ and $A_{1g}(\mathrm{K})+LA(\mathrm{K})$ would be virtually indistinguishable. So we interpret these peaks as combinations of $A_{1g}(\mathrm{M\ or\ K})+LA(\mathrm{M\ or\ K})$ and $E^1_{2g}(\mathrm{M\ or\ K})+LA(\mathrm{M\ or\ K})$. An associated peak is seen at 176 cm$^{-1}$, which is assigned as $A_{1g}(\mathrm{M\ or\ K})-LA(\mathrm{M\ or\ K})$, a Stokes/anti-Stokes combination scattering which rapidly disappears as the temperature is lowered. The relative enhancement of these peaks for the 1.96-eV excitation may be explained in terms of the resonance with excitons. Due to the large exciton binding energy, the wavefunction has a small spatial extent, which means that it contains large-*k* components of the host lattice wave functions. Therefore, two-phonon scattering involving zone-boundary phonons would be relatively enhanced when the excitation energy matches the exciton energy.

**3.5 Resonance enhancement of combination modes for 3.81-eV excitation**

A series of peaks centered at 629, 756, 781, and 822 cm$^{-1}$ are strongly enhanced for the 3.81 eV excitation, especially for 1TL and 2TL. These peaks, except the one at 629 cm$^{-1}$, were reported by Sun *et al*.[45] and were interpreted as being due to 'triply resonant 2-phonon scattering' mediated by the valence bands split by spin-orbit interaction. They proposed that the resonance of the excitation laser energy with the band gap between the top of the valence band and the *c*6 conduction band at the K point explains the strong resonance for the 3.81 eV (325 nm) excitation.45 However, more recent calculations based on GW-Bethe-Salpeter equation (GW-BSE) approach[4] established that the conduction band positions are ~1 eV higher than the values used in Ref. 45. Therefore, the resonance cannot be between the top of the valence band and the *c*6 conduction band, which are separated by ~4.5 eV.[4] Furthermore, the spin-orbit splitting in the valence band for 1TL is 140±10 meV, which corresponds to ~1100 cm$^{-1}$ and does not match the Raman shifts of the peaks. Based on the GW-BSE results,[4] we propose a modified model for these peaks. Since the spin-orbit splitting in the valence band gets smaller as *k* moves away from the K point, one can find a $k$ value for which the spin-orbit splitting matches the energy of 2 phonons. As shown in Figs. 6c–f, several different processes are possible. The momentum of the phonons involved in each of these processes would be different. However, since the phonon dispersion is relatively flat, the involvement of phonons with different momenta would just broaden the Raman peaks. Furthermore, since this resonance occurs away from the K point, the electronic transition energy is somewhat larger than the K-point band gap of ~2.8 eV and close to the excitation energy of 3.81 eV. This would explain why these peaks are strongly enhanced for 3.81 eV and not for 2.81 eV. The conduction bands tend to move down as the number of TLs increases. This also explains why these peaks are most prominent only for 1TL and 2TL and become weaker for thicker layers.

## 4. Conclusions

In summary, we investigated various Raman features of $MoS_2$ thin films as a function of the number of TLs with 6 excitation energies (1.96, 2.33, 2.41, 2.54, 2.81 and 3.81 eV). The resonance behaviour of the two well-known peaks $E^1_{2g}$ and $A_{1g}$ establishes the band gap energy of ~2.8 eV for 1TL and ~2.5 eV for few layer $MoS_2$. In addition, many first- and second-order Raman peaks which depend on the excitation energy are observed. These peaks are explained in terms of resonance with excitons or exciton-polaritons. In the low-frequency range, shear and breathing modes show clear thickness dependence and so may be used as fingerprints of the thickness. Several new features also appear in the low-frequency range for the excitation energy of 1.96 eV, which are related to the resonance with excitons.

## Acknowledgements

This work was supported by the National Research Foundation (NRF) grants funded by the Korean government (MSIP) (Nos. 2011-0013461 and 2011-0017605) and by a grant (No. 2011-0031630) from the Center for Advanced Soft Electronics under the Global Frontier Research Program of MSIP.

## Notes and references

† Electronic Supplementary Information (ESI) available. See DOI: 10.1039/b000000x

‡ Both authors contribute equally to this work.

**SYNOPSIS TOC**

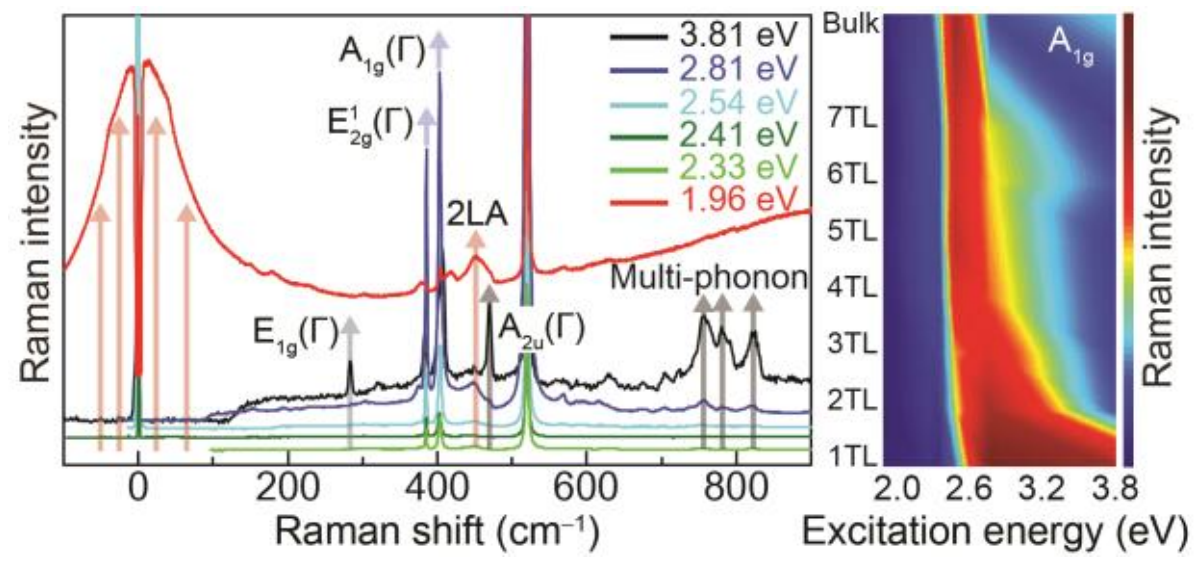

# Electronic Supplementary Information (ESI)

# Anomalous excitonic resonance Raman effects in few-layer $MoS_2$

*Jae-Ung Lee,* [1,†] *Jaesung Park,* [1,†] *Young-Woo Son*[2] *and Hyeonsik Cheong*[1,*]

[1]Department of Physics, Sogang University, Seoul 121-742, Korea

[2]School of Computational Sciences, Korean Institute for Advanced Study, Seoul 130-722, Korea

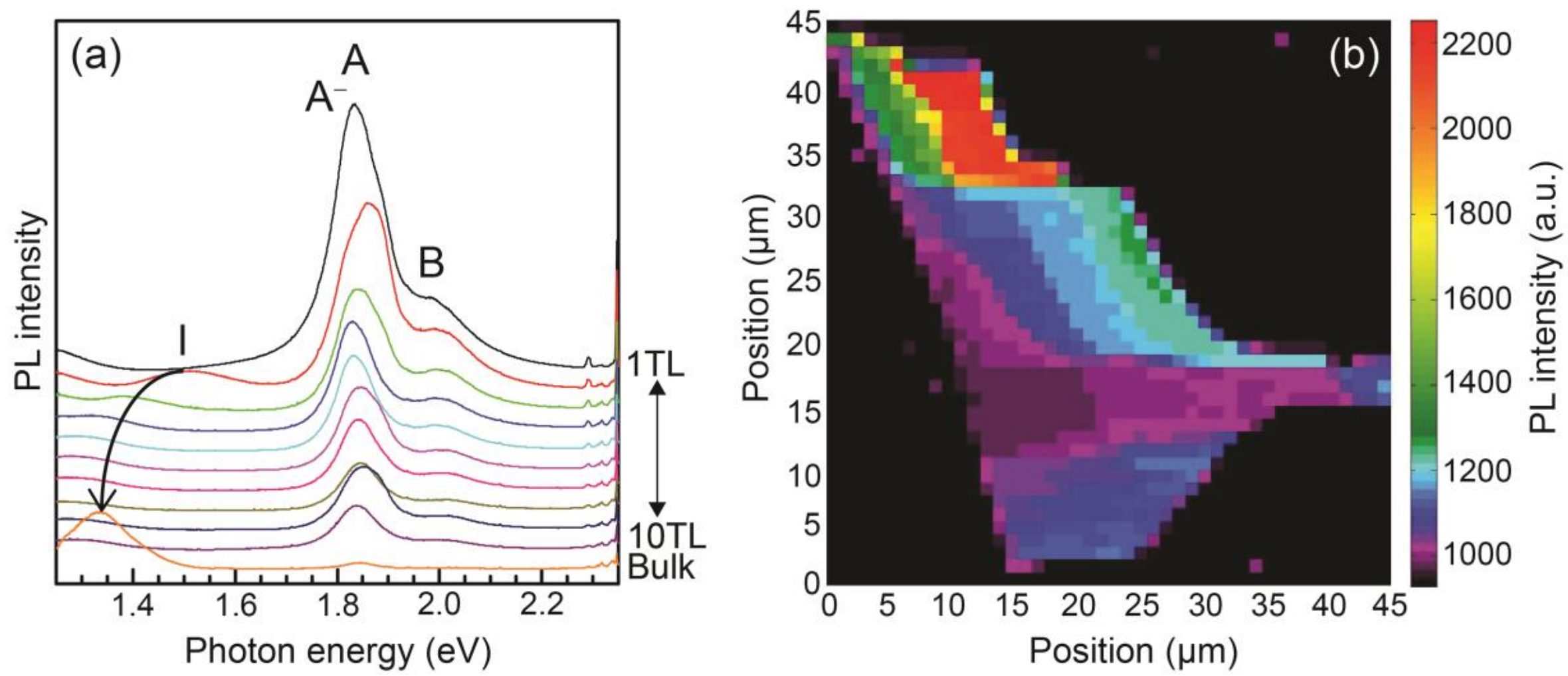

**Figure S1.** (a) Photoluminescence (PL) spectra of few-layer $MoS_2$. (b) PL intensity image of A excition.

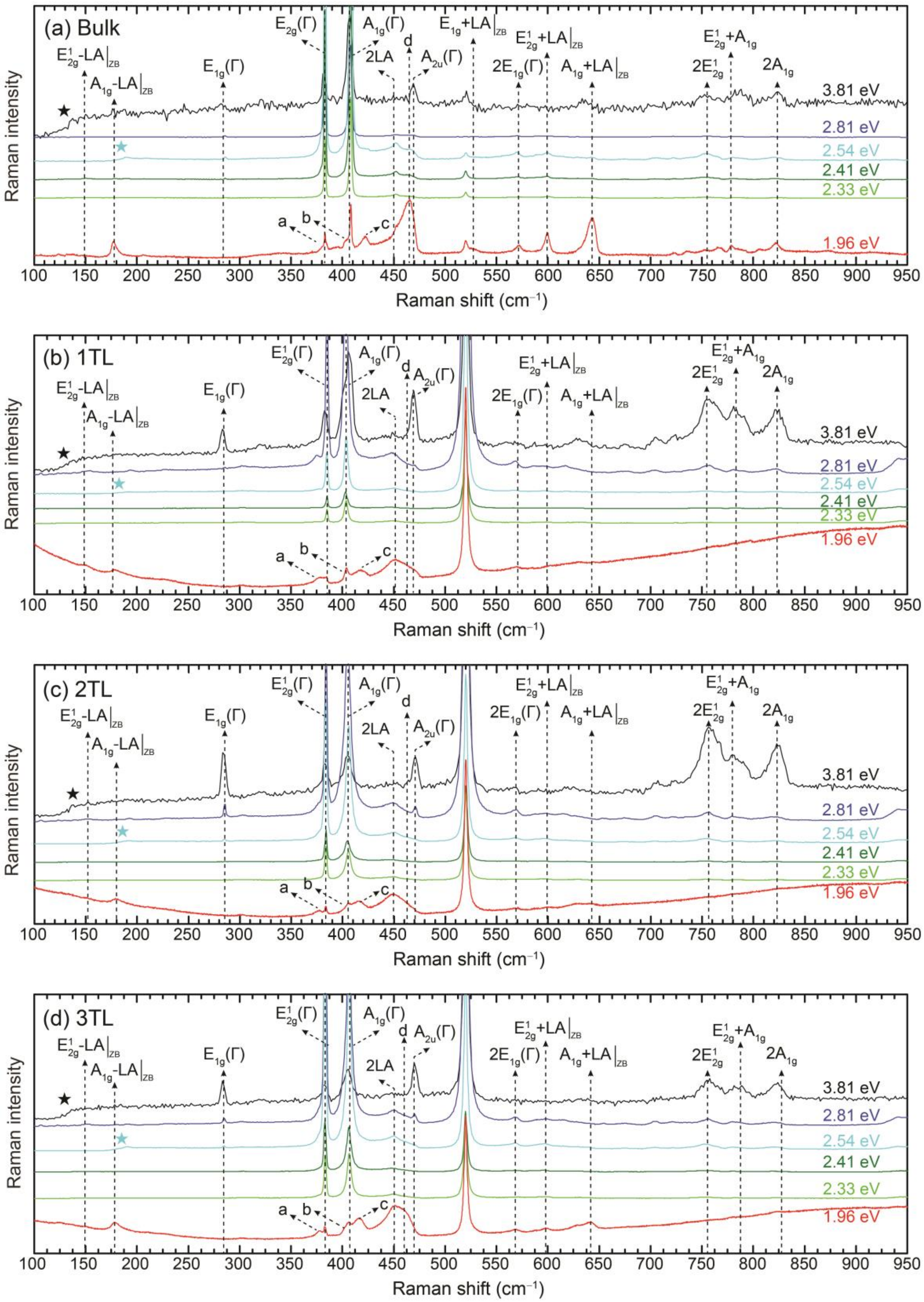

**Figure S2.** Raman spectra of (a) bulk, (b) 1TL, (c) 2TL, and (d) 3TL $MoS_2$ measured with 6 excitation energies: 3.81 eV, 2.81 eV, 2.54 eV, 2.41 eV, 2.33 eV, and 1.96 eV. Features indicated by (★) are experimental artefacts due to the cutoff of the edge filters.

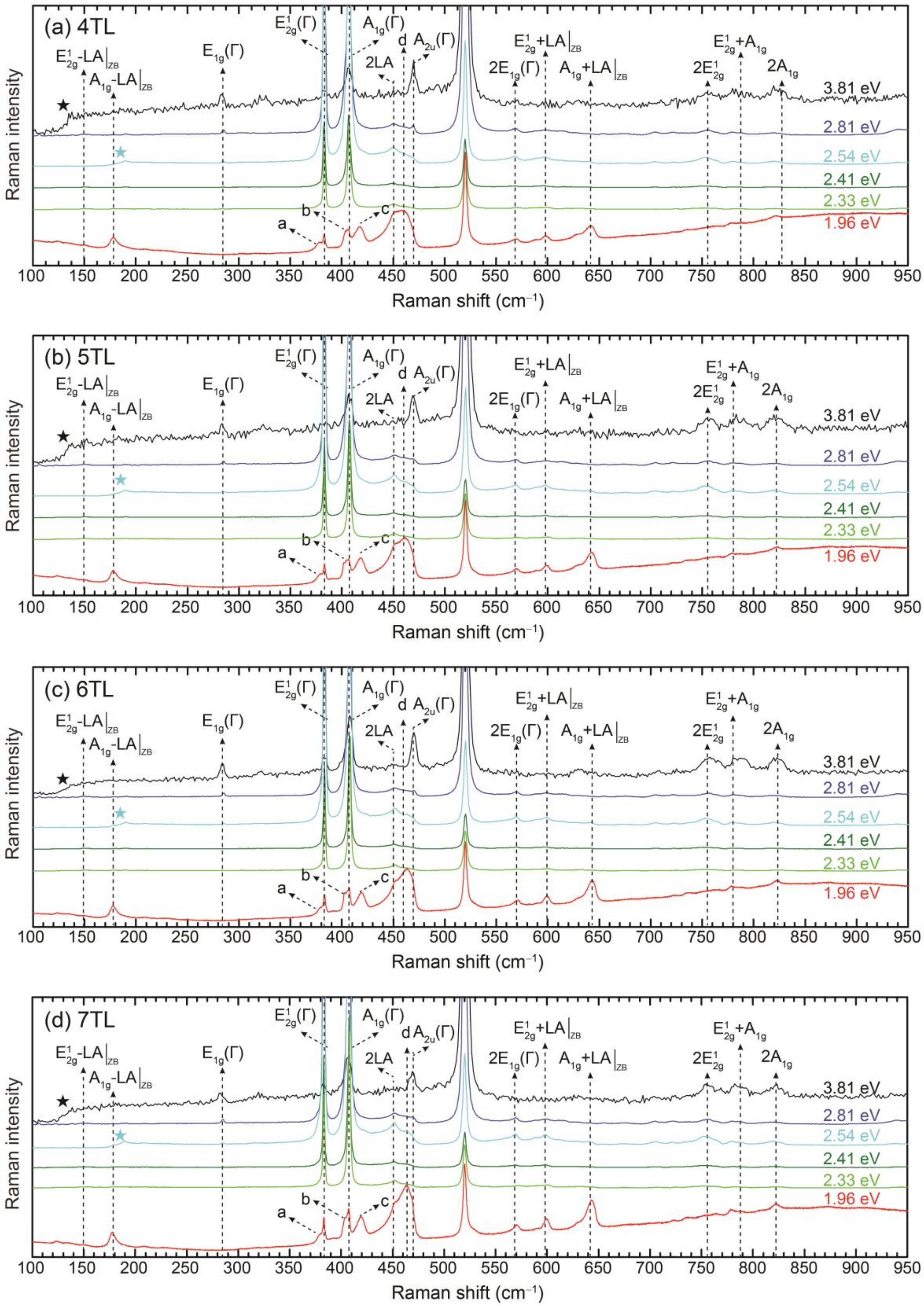

**Figure S3.** Raman spectra of (a) 4TL, (b) 5TL, (c) 6TL, and (d) 7TL $MoS_2$ measured with 6 excitation energies: 3.81 eV, 2.81 eV, 2.54 eV, 2.41 eV, 2.33 eV, and 1.96 eV. Features indicated by (★) are experimental artefacts due to the cutoff of the edge filters.

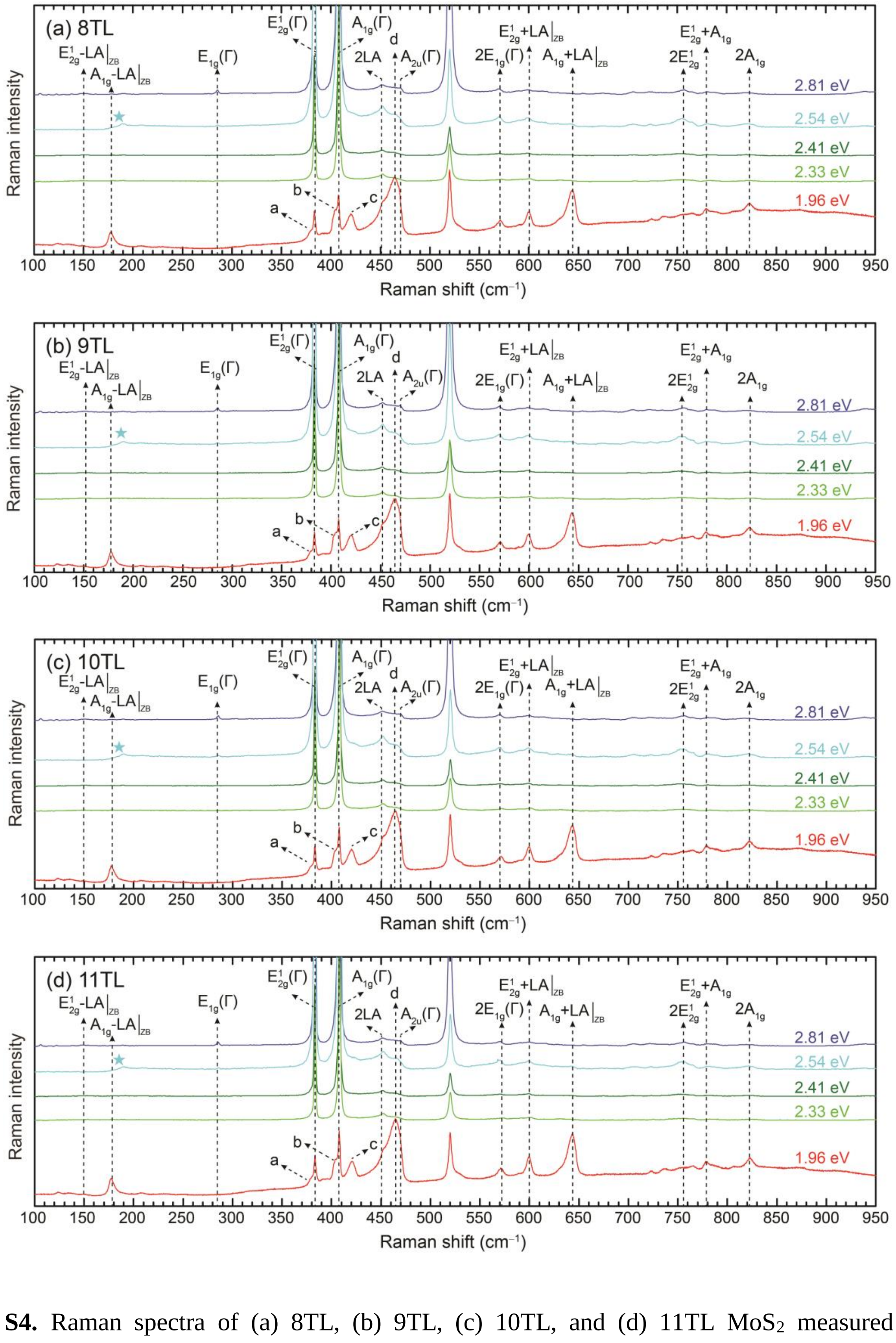

**Figure S4.** Raman spectra of (a) 8TL, (b) 9TL, (c) 10TL, and (d) 11TL $MoS_2$ measured with 5 excitation energies: 2.81 eV, 2.54 eV, 2.41 eV, 2.33 eV, and 1.96 eV. Features indicated by (★) are experimental artefacts due to the cutoff of the edge filters.

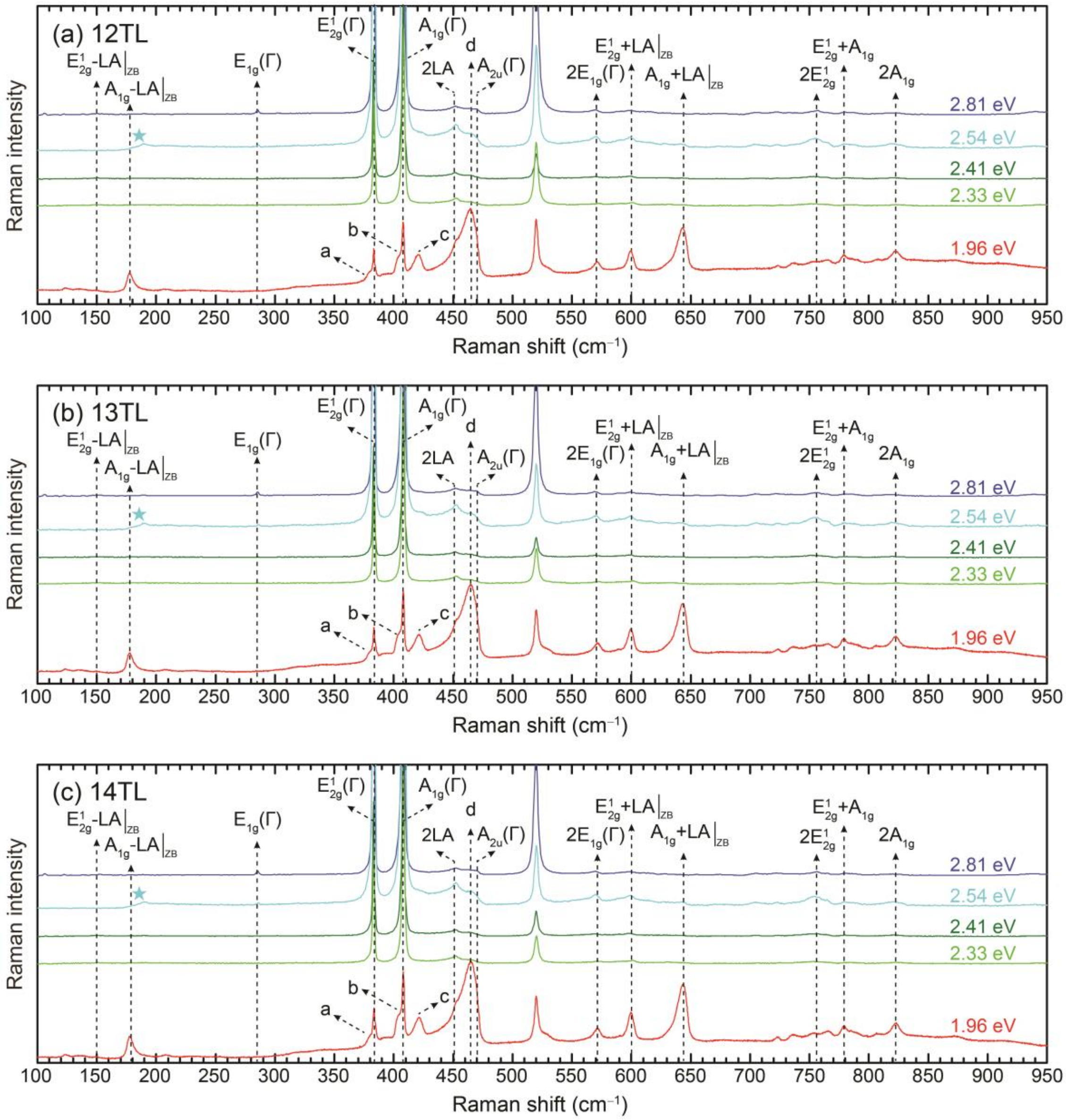

**Figure S5.** Raman spectra of (a) 12TL, (b) 13TL, and (c) 14TL $MoS_2$ measured with 5 excitation energies: 2.81 eV, 2.54 eV, 2.41 eV, 2.33 eV, and 1.96 eV. Features indicated by (★) are experimental artefacts due to the cutoff of the edge filters.

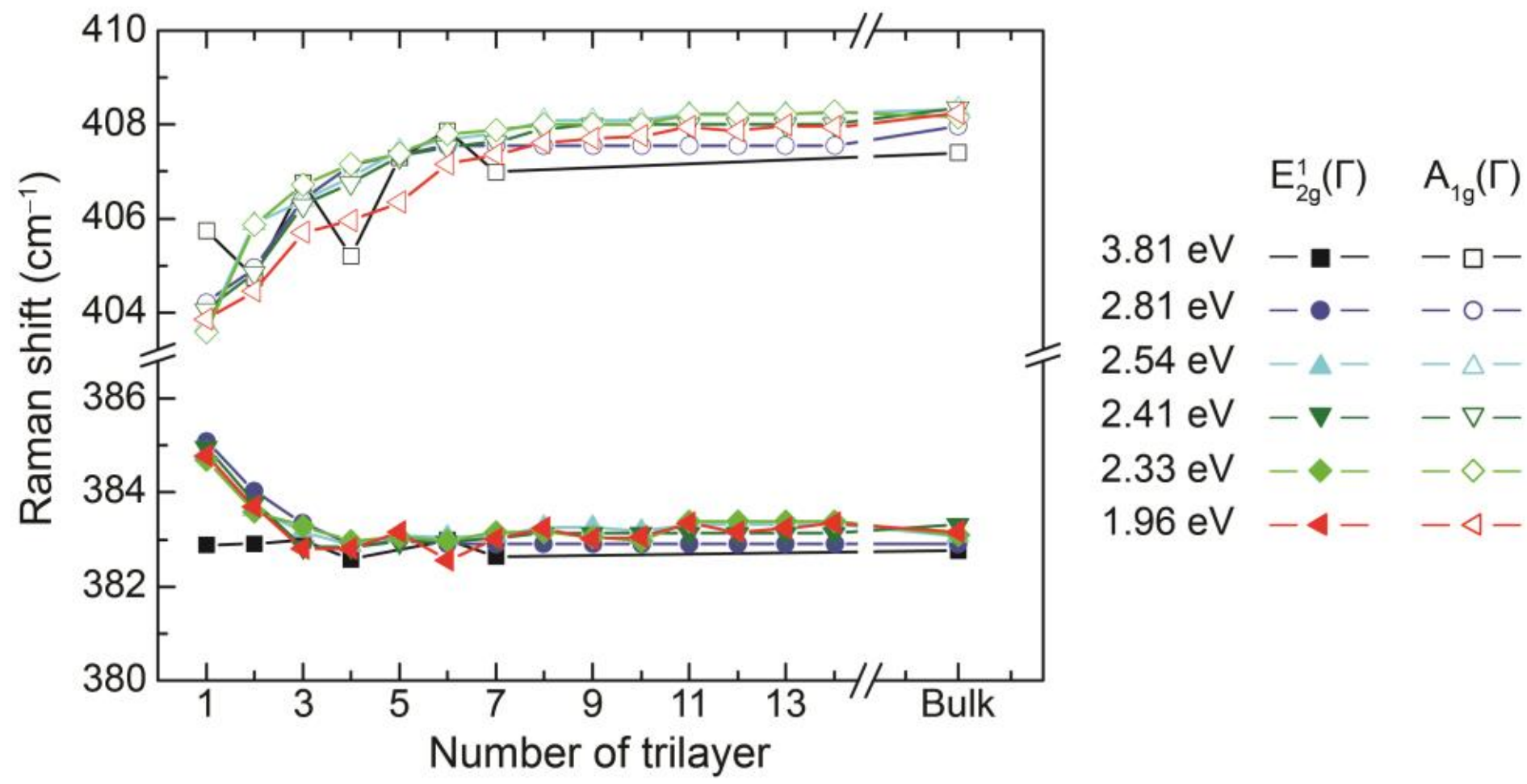

**Figure S6.** Peak positions of $E^1_{2g}$ and $A_{1g}$ modes up to 14TL and bulk for 6 excitation energies. The data for the 3.81 eV excitation show larger scatter due to a lower spectral resolution at short wavelengths.

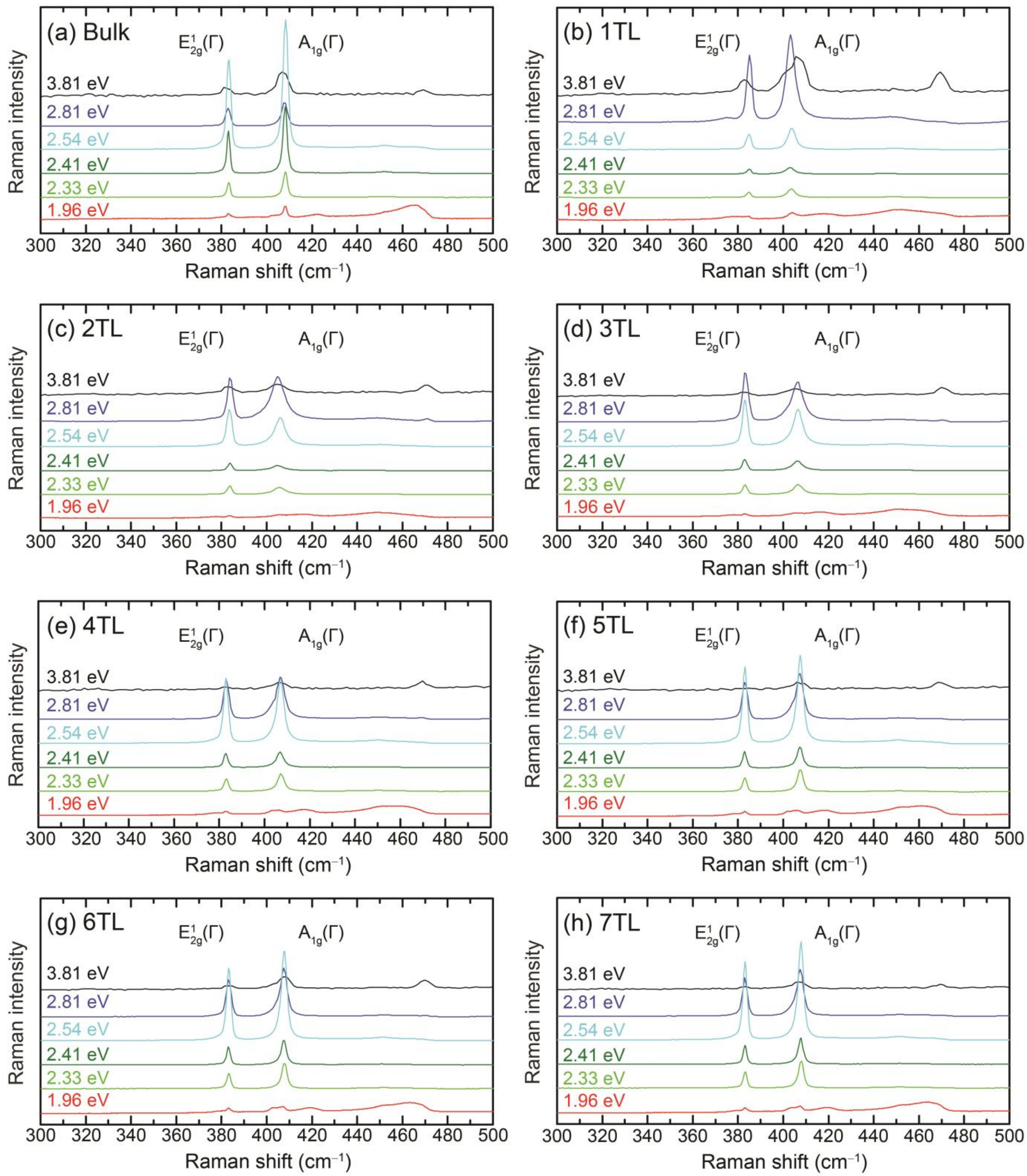

**Figure S7.** Dependence of $E^1_{2g}$ and $A_{1g}$ modes on excitation energies for (a) bulk and (b) 1TL – (h) 7TL $MoS_2$.

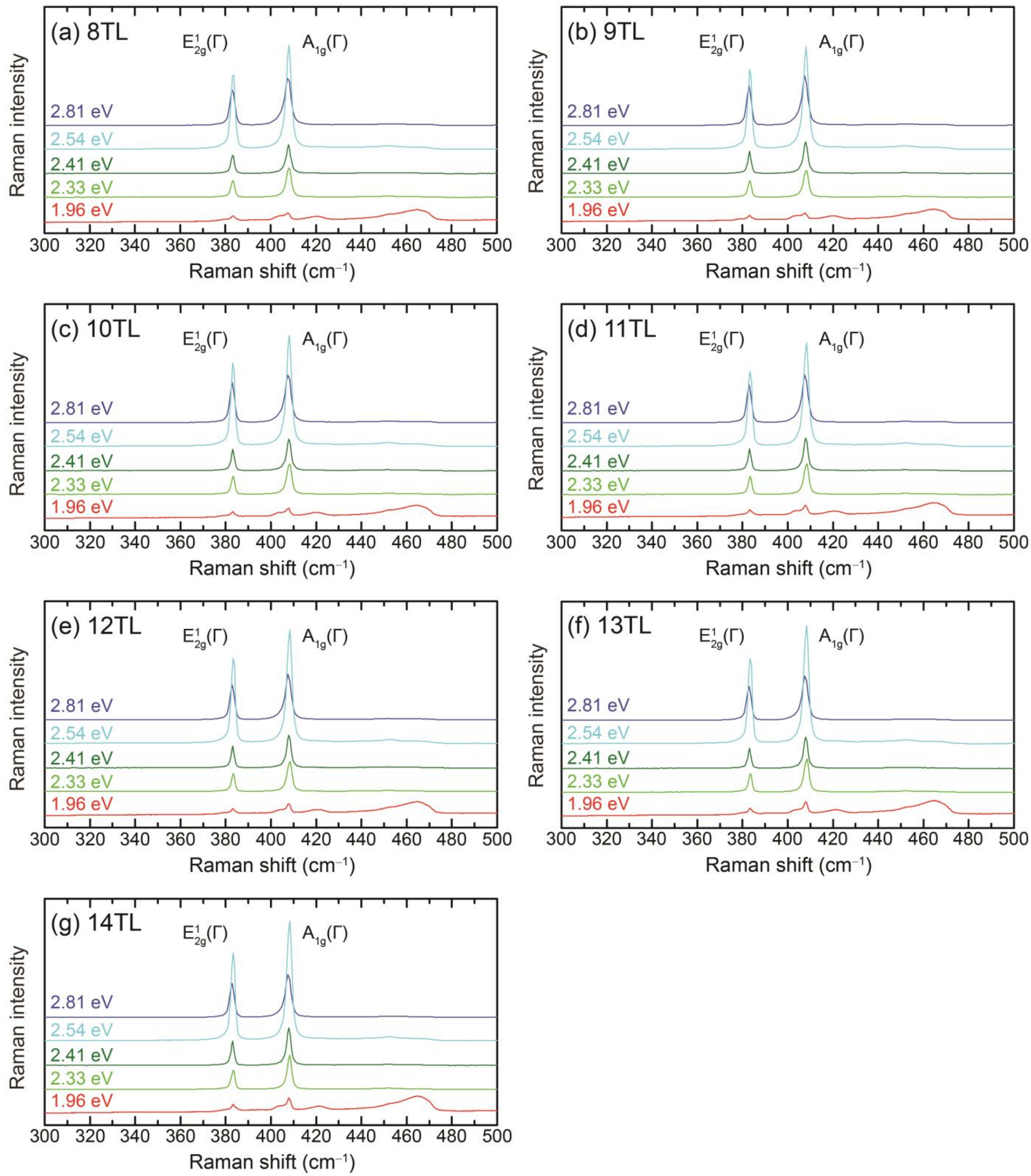

**Figure S8.** Dependence of $E^1_{2g}$ and $A_{1g}$ modes on excitation energies for (a) 8TL – (g) 14TL $MoS_2$.

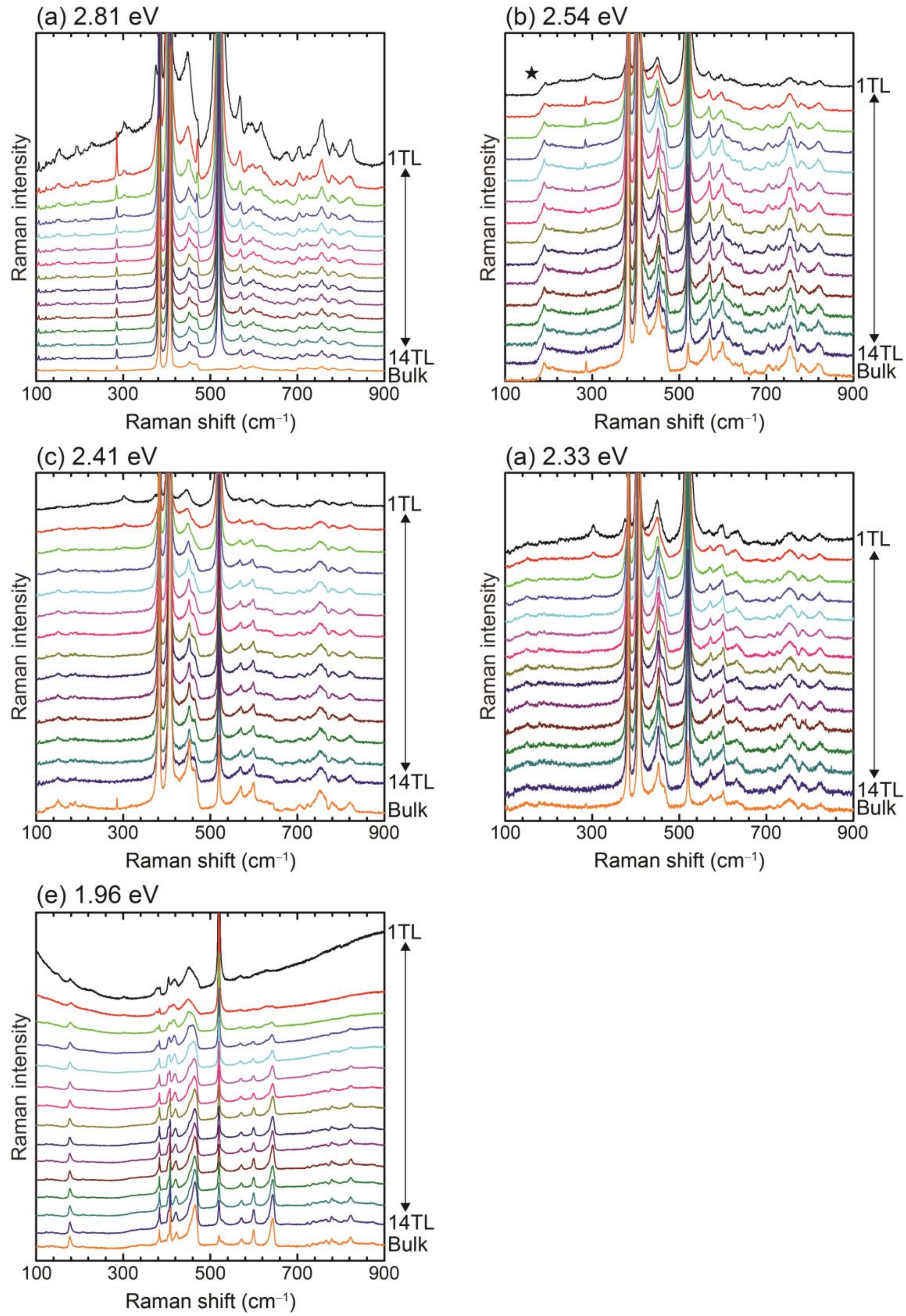

**Figure S9.** Thickness dependence of the Raman spectrum of $MoS_2$ measured with 5 excitation energies: (a) 2.81 eV, (b) 2.54 eV, (c) 2.41 eV, (d) 2.33 eV, and (e) 1.96 eV. Features indicated by (★) are experimental artefacts due to the cutoff of the edge filters.

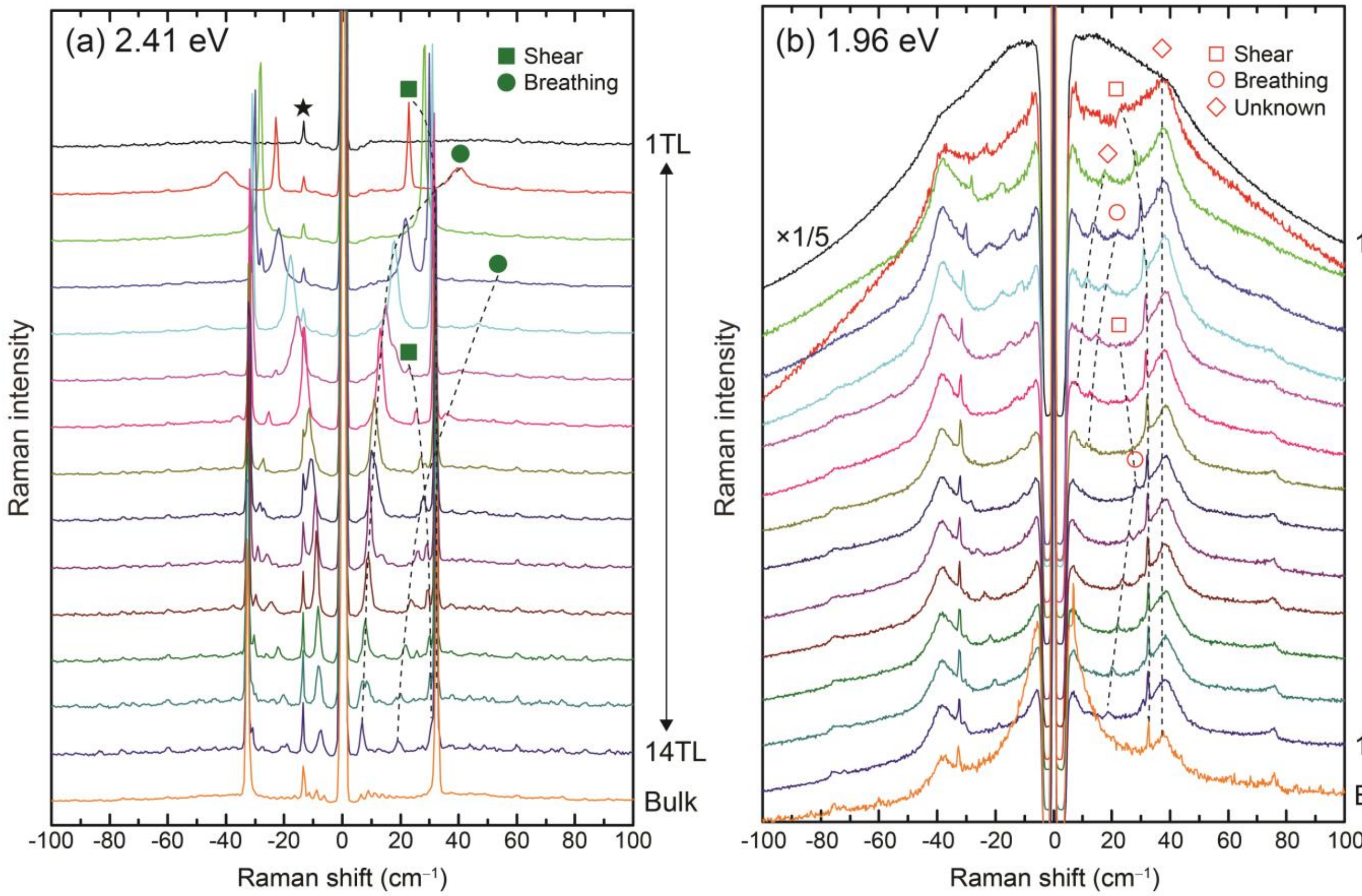

**Figure S10**. Low-frequency Raman spectra of $MoS_2$ measured with excitation energies of (a) 2.41 eV and (b) 1.96 eV. A plasma line of the 2.41-eV laser is indicated by (★).

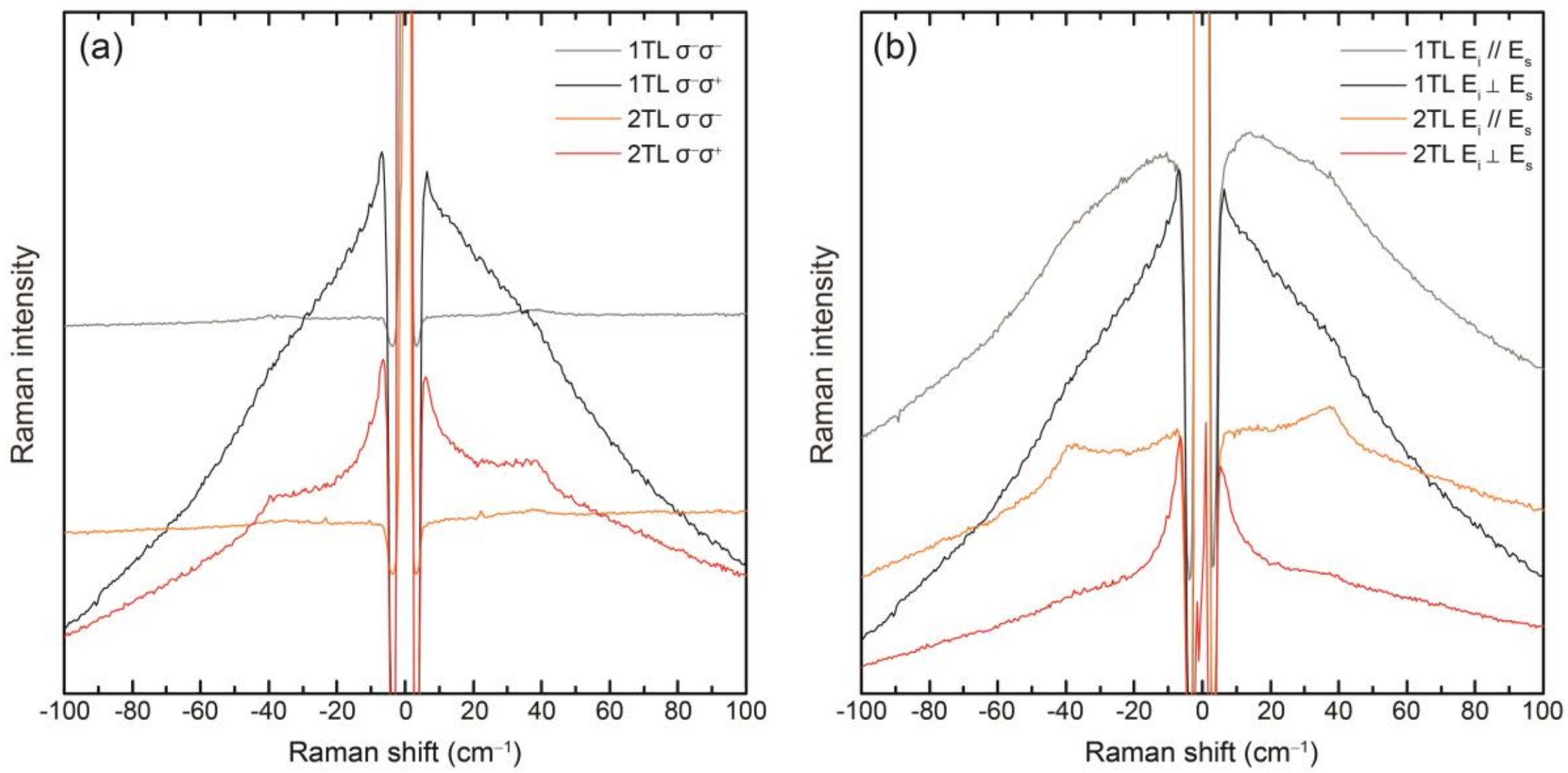

**Figure S11.** Polarization dependence of the low-frequency Raman spectra of 1TL and 2TL $MoS_2$ measured with excitation energy of 1.96 eV: (a) circular and (b) linear polarization dependence. In backscattering geometry, ($\sigma^{-}\sigma^{+}$) or ($\sigma^{+}\sigma^{-}$) correspond to spin-conserving scattering. (a) shows that the 'central peak' is due to a spin-conserving scattering mechanism.

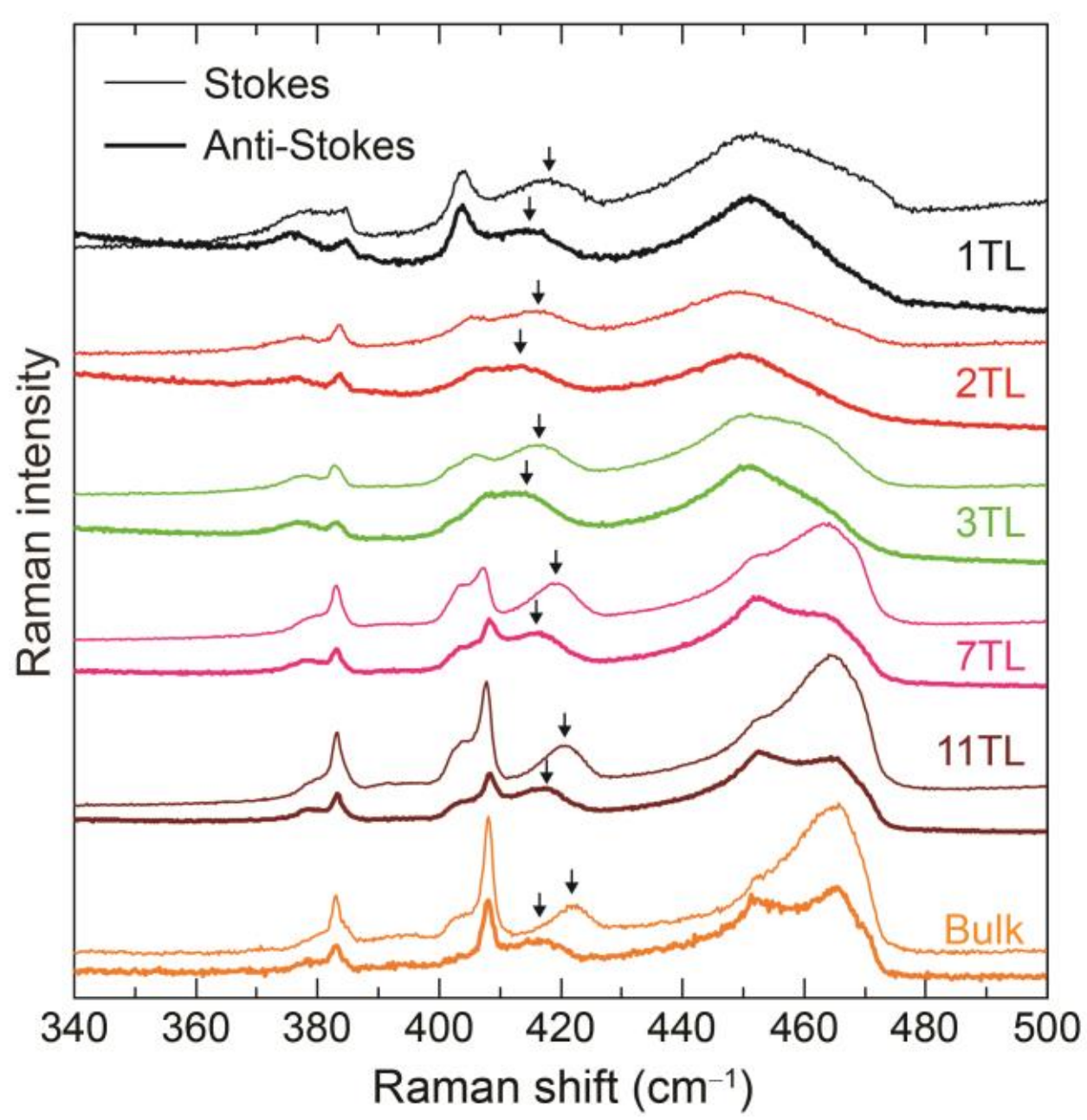

**Figure S12.** Comparison of Stokes and anti-Stokes Raman spectra of few-layer $MoS_2$ for an excitation energy of 1.96 eV. The peak *c* indicated by arrows exhibits a clear displacement between Stokes and anti-Stokes scattering.